\documentclass[twocolumn,aps,amsmath,nofootinbib]{revtex4}

\begin{document}

\title{Probing the minimal length scale by
precision tests of the muon $g-2$}

\author{U.~Harbach}
\email{harbach@th.physik.uni-frankfurt.de}
\author{S.~Hossenfelder}
\author{M.~Bleicher}

\author{H.~St\"ocker}
\affiliation{ \vspace*{3mm}\\
Institut f\"ur Theoretische Physik\\ 
J. W. Goethe-Universit\"at\\
Robert-Mayer-Str. 8-10\\ 
60054 Frankfurt am Main, Germany}

\noindent
\begin{abstract}
Modifications of the gyromagnetic moment of electrons and muons due to a minimal length 
scale combined with a modified fundamental scale $M_f$ are explored.
Deviations from the theoretical Standard Model value for $g-2$ are derived. 
Constraints for the fundamental scale $M_f$ are given.
\vspace{1cm}
\end{abstract}

\maketitle


String theory suggests the existence of a minimum length scale. An exciting quantum mechanical implication of this feature is a modification of the uncertainty principle.  
In perturbative string theory \cite{Gross,Amati:1988tn}, the feature of a fundamental minimal length scale arises from the fact that strings can not probe distances smaller than the string scale. If the energy of a string reaches the Planck mass $m_{\rm p}$, excitations of the string can occur and cause a non-zero extension \cite{witten}. Due to this, uncertainty in position measurement can never become smaller than $l_{\rm p}=\hbar/m_{\rm p}$. For a review, see \cite{Garay:1994en,Kempf:1998gk}.
   
Although a full description of quantum gravity is not yet available, there are general features that seem to go hand in hand with  promising candidates for such a theory: 
\begin{itemize}
\item
the need for a higher dimensional space-time and
\item
the existence of a minimal length scale. 
\end{itemize}

Naturally, this minimum length uncertainty is related to a modification of the standard commutation relations between position and momentum \cite{Kempf:1994su,Kempf:1996nk}. Application of this is of high interest for quantum fluctuations in the early universe and inflation \cite{Hassan,Danielsson:2002kx,Shankaranarayanan:2002ax,Mersini:2001su,Kempf:2000ac,Kempf:2001fa,Martin:2000xs,Easther:2001fz,Brandenberger:2000wr}. The incorporation of the modified commutation relations into quantum theory is not fully consistent in all approaches. We will follow the propositions made in \cite{hossi}. 

In our approach, the existence of a minimal length scale grows important for collider physics at high energies or for high precision measurements at low energies due to the
lowered value of the fundamental scale $M_f$. This new scale is incorporated through the 
central idea of Large eXtra Dimensions ({\sc LXD}s). The model of {\sc LXD}s which was recently proposed in \cite{add1,add2,add3,Randall:1999vf,Randall:1999ee} might indeed 
allow to study first effects of unification or quantum gravity in near future experiments. Here, the hierarchy-problem is solved or at least reformulated in a geometric language by the existence of $d$ compactified {\sc LXD}s in which only gravitons can propagate. The standard model particles are bound to our 4-dimensional sub-manifold, often called our 3-brane. This results in a lowering of the Planck scale to a new fundamental scale $M_f$ and gives rise to the exciting possibility of TeV scale {\sc GUT}s \cite{Dienes}.

In this scenario the following relation between the four-dimensional Planck mass $m_{\rm p}$ and the higher dimensional Planck mass $M_f$ can be derived:
\begin{equation}
m_{\rm p}^2 = R^d M_f^{d+2} \quad.
\end{equation}

The lowered fundamental scale would lead to a vast number of observable phenomena for quantum gravity at energies in the range $M_f$. In fact, the non-observation in past collider experiments of these predicted features gives first constraints on the parameters of the model, the number of extra dimensions $d$ and the fundamental scale $M_f$ \cite{Revcon,Cheung}. On the one hand, this scenario has major consequences for cosmology and astrophysics such as the modification of inflation in the early universe and enhanced supernova-cooling due to graviton emission \cite{add3,Cullen,Probes,astrocon,GOD}. On the other hand, additional processes are expected in high-energy collisions \cite{Rums,Gleisberg:2003ue}: production of real and virtual gravitons \cite{enloss1,enloss2,Hewett,Nussi,Rizzo} and the creation of black holes at energies that can be achieved at colliders in the near future \cite{adm,Giddings:2001ih,Mocioiu:2003gi,Kotwal:2002wg,Uehara:2001yk,Emparan:2001kf,Hossenfelder:2001dn} and in ultra high energetic cosmic rays \cite{Ringwald:2001vk}. One also
has to expect the influence of the extra dimensions on high precision measurements. The
most obvious being the modification of Newton's law at small distances \cite{newton1,newton2,newton3}. Of highest interest are also modifications of the gyromagnetic moment of
Dirac particles which promises new insight into non-Standard Model
couplings and effects \cite{Agashe:2001ra,Cacciapaglia:2001pa,Park:2001xp,Appelquist:2001jz,Nath:2001wf,Calmet:2001si}.

In this paper we study implications of these extensions in the Dirac equation without the aim to derive them from a fully consistent theory. Instead we will analyse possible observable 
modifications that may arise by combining the assumptions of both extra dimensions and a minimal length scale.


In order to implement the notion of a minimal length $L_f$, let us now suppose that one can increase $p$ arbitrarily, but that $k$ has an upper bound. This effect will show up when $p$ approaches a certain scale $M_f$. The physical interpretation of this is that particles could not possess arbitrarily small Compton wavelengths $\lambda = 2\pi/k$ and that arbitrarily small scales could not be resolved anymore. 

To incorporate this behaviour, we assume a relation $k=k(p)$ between $p$ and $k$ which is an uneven function (because of parity) and which asymptotically approaches $1/L_f$. Furthermore, we demand the functional relation between the  energy $E$ and the frequency $\omega$ to be the same as that between the wave vector $k$ and the momentum $p$. A possible choice for the relations is

\begin{eqnarray}
L_f k(p) &=& \tanh^{1/\gamma} \left[ \left( \frac{p}{M_f} 
\right)^{\gamma} \right] \quad, \label{eq1}\\
L_f \omega(E) &=& \tanh^{1/\gamma} \left[ \left( \frac{E}{M_f} \right)^{\gamma} 
\right]\quad,\label{eq2}
\end{eqnarray}
with a real, positive constant $\gamma$.

In the following we will study an approximation, namely the regime of first effects including the orders $(p/M_f)^3$.
For this purpose, we expand the function in a Taylor series for small arguments.
 
Because the exact functional dependence is unknown, we assume an arbitrary factor $\alpha$ in front of the order $(p/M_f)^3$-term. Therefore the most general relations for $k(p)$ and $\omega(E)$ which we will use in the following should be
\begin{eqnarray}
L_f k(p) &\approx& \frac{p}{M_f} - \alpha \left(\frac{p}{M_f}\right)^{3} \label{app1a}\\
L_f \omega(E) &\approx& \frac{E}{M_f} - \alpha \left( \frac{E}{M_f} \right)^{3}\\
\frac{1}{M_f} p(k) &\approx& k L_f +\alpha\left( k L_f \right)^3\\
\frac{1}{M_f} E(\omega) &\approx& \omega L_f+\alpha\left( \omega L_f \right)^3 \label{app4a} \quad,
\end{eqnarray}
with $\alpha$ being of order one, in general negative values of $\alpha$ can not be excluded.

This yields to 3$^{\rm rd}$ order
\begin{eqnarray}
\frac{1}{\hbar} \frac{\partial p}{\partial k} &\approx& 1 + 3\alpha\left(\frac{p}{M_f} \right)^2 \label{diff2a}\quad.
\end{eqnarray}

The quantisation of these relations is straight forward. The commutators between $\hat{k}$ and $\hat{x}$ remain in the standard form:
\begin{eqnarray}
[\hat{x},\hat{k}]= {\rm i}\delta_{ij}\quad.
\end{eqnarray}

Inserting the functional relation between the wave vector and the momentum then yields the modified commutator for the momentum. With the commutator relation  
\begin{eqnarray}
[\,\hat{x}, \hat{A}(k)] = + {\rm i} \frac{\partial A}{\partial k} \quad,
\end{eqnarray}
the modified commutator algebra now reads
\begin{eqnarray}
[\,\hat{x},\hat{p}]= + {\rm i} \frac{\partial p}{\partial k} \quad.
\end{eqnarray}
This results in the generalised uncertainty relation
\begin{eqnarray}
\Delta p \Delta x \geq \frac{1}{2}  \Bigg| \left\langle \frac{\partial p}{\partial k} \right\rangle \Bigg| \quad.
\end{eqnarray}

With the approximations (\ref{app1a})-(\ref{app4a}), the results of Ref. \cite{Hassan} are reproduced up to the factor $\alpha$:  
\begin{eqnarray}
[\hat{x},\hat{p}] \approx {\rm i} \hbar\left( 1 + 3\alpha\frac{ \hat{p}^2}{M_f^2} \right)
\end{eqnarray}
giving the generalised uncertainty relation
\begin{eqnarray}
\Delta p \Delta x \geq \frac{1}{2} \hbar \left( 1+ 3\alpha\frac{\langle \hat{p}^2 \rangle }{M_f^2} \right) \quad.
\end{eqnarray}

Quantisation proceeds in the usual way from the commutation relations. Focusing on conservative potentials in quantum mechanics we give the operators in the position representation which is suited best for this purpose:
\begin{eqnarray}
\hat{x} &=& x\quad,\quad \hat{k}= - {\rm i} \partial_x\nonumber\\
\hat{p} &=& \hat{p}(\hat{k}) \quad,  
\end{eqnarray}
yielding the new momentum operator
\begin{eqnarray}
\hat{p} \approx - {\rm i} \hbar \left( 1 - \alpha L_f^2 \partial_x^2 \right)\partial_x \quad.
\end{eqnarray} 

In ordinary relativistic quantum mechanics the Hamiltonian of the Dirac Particle is \footnote{Greek indices run from 0 to 3, roman indices run from 1 to 3.}
\begin{eqnarray}
\hat{H}= {\rm i} \hbar  \partial_0 = \gamma^0\left( {\rm i} \hbar\gamma^i\partial_i +m \right).
\end{eqnarray}

This leads to the Dirac Equation
\begin{eqnarray}
\label{dirac} (p\hspace{-1.5mm}/-m)\psi = 0 \quad,
\end{eqnarray}
with the following standard abbreviation $\gamma^\nu A_{\nu} := A\hspace{-1.5mm}/$ and $p_{\nu} = {\rm i} \hbar \partial_{\nu}$. To include the modifications due to the generalised uncertainty principle, we start with the relation
\begin{eqnarray}
\hat{E}(\omega)= \gamma^0\left(\gamma^i\hat{p}_i(k) + m \right)
\end{eqnarray}
as the first step to quantisation. Including the altered momentum wave vector relation $\hat{p}(k)$, this yields again Eq. (\ref{dirac}) with the modified momentum operator
\begin{eqnarray}
(p\hspace{-1.5mm}/(k)-m)\psi = 0 \quad.
\end{eqnarray}

This equation is Lorentz invariant by construction. Since it contains -- in position representation -- 3rd order derivatives in space coordinates, it contains 3rd order time-derivatives too. In our approximation, we can solve the equation for a single order time derivative by using the energy condition $E^2=p^2+m^2$. This leads effectively to a replacement of time derivatives by space derivatives:
\begin{eqnarray}
\hbar \hat{\omega} \approx \hat{E} - \alpha\hat{E}^3/M_f^2 = \hat{E}\left(1-\alpha\frac{\hat{p}^i\hat{p}_i + m^2}{M_f^2} \right) \; .
\end{eqnarray} 
Inserting the modified $\hat{E}(\omega)$ and $\hat{p}(k)$ and keeping only derivatives up to 3rd order, we obtain the following expression of the Dirac Equation:
\begin{eqnarray}
\label{dirac2}
\omega \vert \psi \rangle \approx  \gamma^0\left(\gamma^i\hat{k}_i +\frac{m}{\hbar} \right) \left(1-\alpha\frac{\hbar^2 \hat{k}^i\hat{k}_i + m^2}{M_f^2} \right) \vert \psi \rangle \; .
\end{eqnarray} 

The task is now to derive the modifications of the anomalous gyromagnetic moment 
due to the existence of a minimal length. Therefore we assume as usual the particle
is placed inside a  homogeneous and static magnetic field $B$. Regarding the
energy levels of an electron the magnetic field leads to a splitting of the 
energetic degenerated values which is proportional to the magnetic field $B$ and the gyromagnetic moment $g$. Since the energy of the particle in the field is not 
modified (see (\ref{dirac})) there is no modification of the splitting as one might
have expected from the fact that the particles spin is responsible for the anomaly.

However, if we look at the precession of a dipole in a magnetic field without minimal
length and compare its precession frequency to that of the spin $1/2$ particle under
investigation, again the factor $g$ occurs. Without minimal length the frequency
from quantum mechanics is two times the classical one. In that case a further
modification from the minimal length are expected since the
relation between energy and frequency is involved. Thus,  the 
modification of $g$ depends crucially on the way it is
measured. Let us now derive this novel formulation. 

Equation (\ref{dirac2}) with minimally coupled electromagnetic fields reads:
\begin{eqnarray}
\omega \vert\psi\rangle \approx \gamma^0\left( \gamma^i\hat{K}_i +\frac{m}{\hbar}\right)\left(1-\alpha\frac{\hbar^2 \hat{K}^i\hat{K}_i + m^2}{M_f^2}\right)\vert\psi\rangle
\end{eqnarray} 
where $\hat{K}=\hat{k}+ e \hat{A}/\hbar$. Higher derivatives acting on the
magnetic potential can be dropped too for a static and uniform field. In addition, the
constant electric potential can be set to zero. In the non-relativistic approximation 
we can simplify this equation in the Coulomb gauge to:
\begin{eqnarray}
\left( E+m\hat{F}  \right) \vert \chi\rangle =
\left( \frac{(\hbar\hat{K})^2}{2 m}\hat{F}+\frac{e\hbar}{2 m}\sigma\cdot\hat{B}\hat{F}  \right) \vert\chi\rangle
\end{eqnarray} 
with 
\begin{eqnarray}
\hat{F}=\left( 1- \alpha \frac{\hbar^2 \hat{K}^i\hat{K}_i+m^2}{M_f^2} \right) \quad,\quad \vert\psi\rangle = \left\vert{\chi\atop\phi}\right\rangle 
\end{eqnarray} 
Here $\chi$ is the upper component of the Dirac spinor and $\sigma$ denotes the Pauli matrices. 

Therefore, the modified expression $\tilde{g}$ for the gyromagnetic moment is:
\begin{eqnarray}
\tilde{g}=g\cdot\bigg(1-\alpha \frac{m^2}{M_f^2}\bigg)
\end{eqnarray} 

The experimental data concerning the muon gyromagnetic moment are as follows: Davier and collaborates provide two standard model theory results; they differ in the experimental input\footnote{The indices indicate the source of the vector spectral functions; they are obtained by either hadronic $\tau$ decays or $e^+e^-$-annihilation cross-sections.} used to the hadronic contributions \cite{Davier:2002dy}. It is convenient to use the quantity $a = (g-2)/2$:
\begin{eqnarray}
a_{\mu,\tau} &=& 11659193.6(10.9) \times 10^{-10}\nonumber\\
a_{\mu,ee} &=& 11659169.3(9.8) \times 10^{-10}\nonumber
\end{eqnarray}

The experimental 'world average' is \cite{Bennett:2002jb}:
\begin{equation}
a_\mu=11659203(8) \times 10^{-10}\quad.
\end{equation}

The results indicate that modifications to the standard model calculation have to
be smaller than $10^{-8}$. This leads to the following constraint on the fundamental scale
of the theory:
\begin{equation}
M_f/\sqrt{\vert\alpha\vert} \ge 1~{\rm TeV}\quad.
\end{equation}

As we are working within a model with large extra dimensions, there might further be corrections due to graviton loops \cite{Graesser:1999yg,Kim:2001rc}. However, recent calculations show that neither sign nor value of these corrections are predictable due to unknown form-factors and cutoff parameters \cite{Contino:2001nj}.


A model, which combines both Large Extra Dimensions and the minimal length scale $L_f$ is studied.
The existence of a minimal length scale leads to modifications of quantum mechanics. 
With the recently proposed idea of Large Extra Dimensions, this new scale might be in reach
of present day experiments. The modified Dirac equation is used to derive 
an expression for the gyromagnetic moment of  spin $1/2$ particles. Our results for the muon $g-2$ value are compared to the values predicted by QED and experiment. For $\gamma=1$ ($\alpha=1/3$), a specific limit on the fundamental scale $M_f$ can be obtained from present $g-2$ data: $M_f \ge  577$~GeV. 

\section*{Acknowledgements}

The authors thank  S.~Hofmann and J. Ruppert for fruitful discussions. 
S.~Hossenfelder wants to thank the Land Hessen for financial support.

\end{document}